\documentstyle[11pt]{article}
\def\cF{\cal F}

\newfont{\goth}{eufm10 scaled \magstep1}

\def\a{{\alpha}}

\def\c{\gamma}

\def\k{\kappa}
\def\L{\Lambda}

\def\S{\Sigma}

\def\o{\omega}
\def\del{\partial}
\def\ua{\underline{\alpha}}

\def\una{\underline a}\def\unA{\underline A}

\def\unM{\underline M}

\def\atwo{\alpha_{2}}
\def\aone{\alpha_{1}}
\def\afive{\alpha_{5}}
\def\ap{\alpha_p}

\def\appt{\alpha_{p+2}}
\def\apmo{\alpha_{p-1}}

\def\a3t{   {a_3}_{\parallel}   }
\def\a3n{{a_3}_{\perp}}
\def\a#1{{\alpha_#1}} 

\let\la=\label

\let\bm=\bibitem

\def\unM{{\underline M}}

\def\bd{\begin{document}}
\def\ed{\end{document}}
\def\ba{\begin{array}}
\def\ea{\end{array}}
\def\bea{\begin{eqnarray}}
\def\eea{\end{eqnarray}}
\def\ft#1#2{{\textstyle{{\scriptstyle #1}\over {\scriptstyle #2}}}}
\def\fft#1#2{{#1 \over #2}}
\newcommand{\be}{\begin{equation}}
\newcommand{\ee}{\end{equation}}
\newcommand{\eq}[1]{(\ref{#1})}
\def\eqs#1#2{(\ref{#1}-\ref{#2})}
\def\det{{\rm det\,}}
\def\tr{{\rm tr}}
\newcommand{\ho}[1]{$\, ^{#1}$}
\newcommand{\hoch}[1]{$\, ^{#1}$}
\def\ra{\rightarrow}
\def\uha{{\hat {\underline{\a}} }}
\def\uhc{{\hat {\underline{\c}} }}

\newcommand{\tamphys}{\it\small Center for Theoretical Physics, Texas
A\&M University, College Station, TX 77843, USA}

\newcommand{\sissa}{\it\small International School for Advanced Studies
(SISSA/ISAS), Via Beirut 2, 34014 Trieste, Italy}

\newcommand{\kings}{\it\small Department of Mathematics, King's College,
London, UK}

\newcommand{\auth}{\large C.S. Chu\hoch{1}, P.S. Howe\hoch{2}, 
E. Sezgin\hoch{3\dagger} and P.C. West\hoch{2}} 

\begin{document}

\hfill{SISSA 16/98/FM}

\hfill{KCL-MTH-98-06}

\hfill{CTP TAMU-5/98}

\hfill{hep-th/9803041}

\vspace{20pt}

\begin{center}

{\Large\bf Open Superbranes}
\vspace{30pt}

\auth

\vspace{15pt}

\begin{itemize}
\item[$^1$] {\small \em
International School for Advanced Studies (SISSA),
Via Beirut 2, 34014 Trieste, Italy}
\item[$^2$] {\small \em
Department of Mathematics, King's College, London, UK}
\item[$^3$] {\small \em 
Center for Theoretical Physics, Texas A\&M University, 
College Station, TX 77843, USA}
\end{itemize}

\vspace{60pt}

{\bf Abstract}

\end{center}

Open branes ending on other branes, which may be referred to as the host
branes, are studied in the superembedding formalism. The open brane,
host brane and the target space in which they are both embedded are all
taken to be supermanifolds. It is shown that the superspace constraints
satisfied by the open brane are sufficient to determine the
corresponding superspace constraints for the host branes, whose dynamics
are determined by these constraints. As a byproduct, one also obtains
information about the boundary of the open brane propagating in the host
brane. 

{\vfill\leftline{} \vfill
\vskip	10pt
\footnoterule
{\footnotesize \hoch{\dagger} Research supported in part by NSF Grant
PHY-9722090 \vskip -12pt}
\vskip	10pt

\pagebreak
\setcounter{page}{1}

\section{Introduction}

It is well-known that certain open branes can end on other branes
provided that the dimensionalities of the branes are chosen correctly
[1-16]. These configurations are intimately connected with the special
cases of intersecting branes where the dimension of the intersection
manifold coincides with one of the intersecting brane boundary
dimension. This is known to occur for fundamental string or
D$(p-2)$-branes ending on D$p$-branes and D$p$-branes ending on NS
5-branes ($1\le p\le 6$). (See \cite{ob} and \cite{bgt} for an extensive
analysis). 

Recently, such systems have been discussed in a hybrid Green-Schwarz
formalism in M-theory and in the context of D-branes in ten dimensions
\cite{cs,chs}. (For earlier related work, see
\cite{mc,brax1,ezawa,brax2}). The basic idea of references \cite{cs} and
\cite{chs} is to write down the Green-Schwarz (GS)-action for a brane,
for example the M$2$-brane, with a boundary. The worldvolume for the GS
action is a bosonic manifold $\S$ while the target space in this case is
the superspace for eleven-dimensional supergravity, $\unM$. The boundary
of the worldvolume, $\del\S$, is taken to be embedded in a supermanifold
$M$ which is also embedded in $\unM$. The supermanifold $M$ is taken to
be the worldsurface of the M$5$-brane, and it was found that
$\k$-symmetry of the M$2$-brane action with boundary contained in $M$
imply the equations of motion for the M$5$-brane. These results were
extended to strings and D$(p-2)$ branes ending on D$p$-branes in
\cite{chs}. The formalism used in these papers is therefore a hybrid
one, since the first brane is treated from the GS point of view while
the second one is treated using the superembedding formalism. Since
$\k$-symmetry is a relic of local worldsurface supersymmetry in the
superembedding formalism it is to be expected that the same results can
be obtained by working entirely in the superembedding formalism. This is
indeed the case as we shall show in the current paper. We shall again
focus on M-branes and D-branes. 

In the approach presented here, it is not necessary to exhibit the
open-host brane system as a classical solution of the target space
theory, nor is it necessary to make assumption on the topology of the
host brane. For example, the host brane itself may be closed or open. An
open host brane, can in turn end on another suitable secondary host
brane and so on. Thus one can obtain a brain chain. A special case of
this arises when all members of the chain have the same dimension, thus
forming a brane network. The two possibilities allowed are those which
use the string ending on D1-brane junction as building block
\cite{js,ah,as}, and those which use D5-branes ending on NS 5-branes as
building blocks \cite{ah2}. Here, for the purposes of the present paper, we
shall focus on the basic building blocks of chain or network
configurations, namely an open brane ending on a host brane.

The next three sections are devoted to the discussion of an M$2$-brane ending
on an M$5$-brane, fundamental strings ending on D$p$-branes and D$p$-branes
ending on D$(p+2)$-branes, respectively. A summary of our results and
further comments about them are provided in the Conclusions. 

\section{Open branes in M-theory}

We consider the following picture: a 2-brane worldsurface $M_2$, with
(even$|$odd) dimension $(3|16)$ can end on an M$5$-brane worldsurface
$M_5$ (dimension $(6|16)$), via a boundary $M_1$ which is a
supermanifold of dimension $(2|8)$. Thus $M_1= \del M_2$ and $M_1\subset
M_5$ while both $M_2$ and $M_5$ are embedded in an $(11|32)$-dimensional
target space $\unM$. We therefore have embeddings

\be
f_i:M_i \hookrightarrow \unM\ ;\qquad i=1,2,5
\la{1}
\ee

as well as an embedding 

\be
f_1{}^5: M_1\hookrightarrow M_5\ .
\la{2}
\ee

Clearly

\be
f_1=f_5\circ f_1{}^5\ .
\la{3}
\ee

The fact that the 2-brane can end on a supermanifold which has bosonic
dimension two is related to the fact that the 5-brane admits stringlike
soliton solutions to its equations of motion \cite{hlw1}. We shall
comment on the possibility of host branes other than the M$5$-brane in the
Conclusions. 

The boundary of the open membrane, in general, may consist of an
arbitrary number of closed strings. However, to keep the discussion as
simple as possible, we shall consider an open membrane that has the
topology of a disk, and hence a single component boundary, a closed
string. Our analysis can easily be extended for multi-component
boundaries, with essentially same results. 

We shall now demonstrate that if the standard embedding condition is
assumed for the 2-brane then the standard embedding condition for the
5-brane is implied. We shall also show, although it is not essential for
the derivation of the M$5$-brane equations of motion, that this picture
requires for its consistency a 2-form gauge potential $A$ on $M_5$ whose
(modified) 3-form field strength $\cF$ is only non-vanishing when all of
its indices are bosonic. This is not essential for the derivations of
the M$5$-brane equations of motion because, as has been shown elsewhere
\cite{hs2}, this result actually follows from the embedding
condition. However, it is useful to introduce this discussion here as it
will play a more significant r\^{o}le in the analysis of D-branes.

To make the analysis of the embedding conditions we introduce the
embedding matrices $E$ which are simply the derivatives of the
embeddings given above referred to standard bases. Thus we have the
following set of embedding matrices,

\be
E_{A_1}{}^{\unA},\ E_{A_2}{}^{\unA},\ E_{A_5}{}^{\unA}\ \ 
{\rm and}\ \  E_{A_1}{}^{A_5}\ ,
\la{4}
\ee

corresponding to the derivatives of the embeddings $f_1,\ f_2,\ f_5,$ 
and $f_1{}^5$ respectively. The notation here is that underlined 
indices refer to the target space $\unM$ while indices for each of the
manifolds $M_i,\ i=1,2,5$ are distinguished by appending to them the
corresponding numerical subscripts. As usual, indices from the beginning
of the alphabet are preferred basis indices while indices from the
middle of the alphabet denote coordinate indices. Capital indices run
over both bosonic and fermionic indices while latin (greek) letters are
used for bosonic (fermionic) indices separately, for example 
$ A=\left( a,\alpha \right)$.
We shall denote normal indices by primes; it should be clear from the
context which embedding is being referred to when a normal index is
employed. A more explict formula for the embedding matrix is, using the
2-brane as an example, 

\be
E_{A_2}{}^{\unA}=E_{A_2}{}^{M_2} \del_{M_2} z^{\unM} E_{\unM}{}^{\unA}\ ,
\la{5}
\ee

where we have introduced the vielbein matrices, $E_{\unM}{}^{\unA}$ etc., 
which relate the preferred bases to the cooordinate bases.

The basic embedding condition for the 2-brane is

\be
E_{\atwo}{}^{\una}=0\ .
\la{6}
\ee

This equation simply states that the odd tangent space to $M_2$ is a
subspace of the odd tangent space to $\unM$ at each point in
$M_2\subset\unM$. It can be shown that this condition determines the
equations of motion of the membrane and also that the geometry of the
target space is required to be that of on-shell eleven-dimensional
supergravity. On the boundary $M_1=\del M_2$ we therefore have

\be
E_{\aone}{}^{\una} =0\ .
\la{7}
\ee

By the chain rule, we have

\be
E_{A_1}{}^{\unA}=E_{A_1}{}^{A_5} E_{A_5}{}^{\unA}
\la{8}
\ee

and so

\be
0=E_{\aone}{}^{\una}=E_{\aone}{}^{\afive} E_{\afive}{}^{\una} + 
E_{\aone}{}^{a_5} E_{a_5}{}^{\una}\ .
\la{9}
\ee

We now introduce a complementary normal matrix in the bosonic sector for
$M_5$ in $\unM$ denoted by $E_{a'_5}{}^{\una}$. The inverse of the pair
$(E_{a_5}{}^{\una},\,E_{a'_5}{}^{\una})$ is denoted by
$((E^{-1})_{\una}{}^{a_5},\,(E^{-1})_{\una}{}^{a'_5})$. Multiplying
\eq{9} by $(E^{-1})_{\una}{}^{a'_5}$ we find

\be
E_{\aone}{}^{\afive} E_{\afive}{}^{\una} (E^{-1})_{\una}{}^{a'_5}=0\ ;
\la{10}
\ee

while multiplying the same equation by $(E^{-1})_{\una}{}^{a_5}$ we get

\be
E_{\aone}{}^{\afive} E_{\afive}{}^{\una} (E^{-1})_{\una}{}^{a_5} +
E_{\aone}{}^{a_5}=0\ .
\la{11}
\ee

Now for any superembedding it is always possible to choose the odd
tangent space of the embedded submanifold such that (taking $M_5 \subset
\unM$ as an example)

\be
E_{\afive}{}^{\una} (E^{-1})_{\una}{}^{a_5} =0\ .
\la{12}
\ee

To see this we observe that the odd tangent space basis
$E_{\afive}$ for $M_5$ can be written quite generally as

\be
E_{\afive}=E_{\afive}{}^{\ua}E_{\ua} + E_{\afive}{}^{\una} E_{\una}
\la{13}
\ee

while for the even subspace we can write

\be
E_{a_5}=E_{a_5}{}^{\ua}E_{\ua} + E_{a_5}{}^{\una} E_{\una}
\la{14}
\ee

Hence if we redefine $E_{\afive}$ by

\be
E_{\afive}\rightarrow \tilde E_{\afive} = E_{\afive} + 
\L_{\afive}{}^{a_5} E_{a_5}
\la{15}
\ee

for some fermionic superfield $\L_{\afive}{}^{a_5}$ we find

\be
\tilde E_{\afive}{}^{\una} = E_{\afive}{}^{\una}  + 
\L_{\afive}{}^{a_5} E_{a_5}{}^{\una}\ .
\la{16}
\ee

Multiplying this equation with $(E^{-1})_{\una}{}^{a_5}$, we observe
that the quantity $E_{\afive}{}^{\una} (E^{-1})_{\una}{}^{a_5}$ can
always be made to vanish by choosing $\L_{\afive}{}^{a_5}$
aprropriately. Thus, the result \eq{12} is proved. Using this result in
conjunction with \eq{10}, $M_1$ being arbitrary, we see that the odd
tangent space of $M_5$ can always be chosen such that

\be
E_{\afive}{}^{\una}=0
\la{17}
\ee

and this implies, from \eq{11}, that

\be
E_{\aone}{}^{a_5}=0
\la{18}
\ee

as well.

Equation (17) is the standard embedding condition for the M$5$-brane. It
has been shown that this equation determines the equations of motion for
the 5-brane \cite{hs2,hsw1}. Furthermore, the boundary brane $M_1$ also
obeys the standard embedding condition as a subsupermanifold of $M_5$.
We therefore conclude that the 5-brane equations of motion are implied
by requiring the consistency of 2-branes ending on 5-branes.

The embedding constraint \eq{18} for the boundary string confined to
propagate within the 5-brane is rather interesting. We shall comment
further on this point in the Conclusions, but here we shall focus
on the derivation of the 5-brane equations of motion.

Let us now consider the Wess-Zumino form $W_4$ for the 2-brane. As this
brane is type I, i.e. its worldsurface multipet contains only scalars as
bosonic degrees of freedom, $W_4$ is simply the pull-back of the
target-space 4-form field strength of eleven-dimensional supergravity

\be
G_4=d C_3\ . 
\ee

Since $W_4$ is a closed form on a supermanifold with bosonic dimension
three, it must be exact because the de Rham cohomology of a
supermanifold coincides with that of its body. Thus we can write 
\cite{pt,bs,hrs2}

\be
W_4=f_2^*G_4=d K_3
\la{19}
\ee

for some globally defined 3-form $K_3$ on $M_2$. We therefore have

\be
K_3 =d Y_2 + f_2^* C_3\qquad{\rm on} \ M_2\ .
\la{20}
\ee

Since both of the potentials $Y_2$ and $C_3$ are, in general, locally
defined, the fact that $K_3$ is globally defined dictates the
transformation rule for $Y_2$. On the boundary $M_1$ we identify $Y_2$
as the pull-back of a 2-form potential on $M_5$, 

\be
Y_2=(f_1{}^5)^* A_2\qquad{\rm on} \ M_1\ .
\la{21}
\ee

This implies that the corresponding field strength 3-form on $M_5$
should be defined by

\be
{\cF}_3=d A_2 + f_5^* C_3\ \qquad{\rm on} \ M_5\ .
\la{22}
\ee

The associated Bianchi identity is

\be
d {\cF}_3= f_5^* G_4\ .
\la{23}
\ee

It is  straightforward to demonstrate that the only non-vanishing
component of $K_3$ is the one with purely bosonic indices, so that it
must vanish when restricted to the boundary $M_1$. This implies in turn
that ${\cF}_3$ must vanish on $M_1\subset M_5$. The pull-back of
${\cF}_3$ to $M_1$ is given by

\be
(f_1{}^5)^*{\cF}_{A_1 B_1 C_1} = 
E_{C_1}{}^{C_5}E_{B_1}{}^{B_5}E_{A_1}{}^{A_5} {\cF}_{A_5 B_5 C_5}\ ,
\la{24}
\ee

up to Grassmann sign factors which we suppress. Since the left-hand side
vanishes, and since $E_{\aone}{}^{a_5}=0$, we conclude, $M_1$ being
arbitrary, that this can only be satisfied if

\be
{\cF}_{\afive B_5 C_5}=0\ ,
\la{25}
\ee

that is, ${\cF}_3$ must be purely even on $M_5$.

To summarise then, we have shown that the consistency of the picture of
a 2-brane with a boundary ending on a 5-brane in which the boundary is
embedded requires that, if the standard embedding condition \eq{6} for
the 2-brane is imposed, the standard embedding condition for the 5-brane
should also hold and that there is a 2-form gauge potential on $M_5$
whose 3-form field strength should satisfy equation \eq{25} above.
This is in perfect agreement with the results of \cite{cs} obtained from
a hybrid approach mentioned in the introduction. As we have remarked
earlier, the second of these equations actually follows from the first
in the case of M-branes, but this is not true for all D-branes
\cite{hrss}. 

\section{Fundamental strings ending on D-branes}

The discussion of the previous section can be carried over
straightforwardly to fundamental type II strings in ten dimensions
ending on D-branes. In general, the end points of the open string may
lie on two different D$p$-branes or one end-point of a semi-infinite
open string may lie on a D$p$-brane while the other end is feely moving.
It is sufficent to consider the case where both end-points are ending on
a single D$p$-brane for the purpose of deriving the constraints that
govern the dynamics of the D$p$-brane. It is straightforward to
generalize the discussion for the other two cases. Thus we have the
following supermanifolds: the string manifold, $M_1$ (dimension
$(2|16)$), its boundary $M_0=\del M_1$ (dimension $(1|8))$, the
worldvolume of the D$p$-brane $M_p$ (dimension $(p+1|16)$) and the
target space $\unM$ which is either type IIA or type IIB superspace and
which has dimension $(10|32)$. The associated embeddings are

\be
f_i:M_i\hookrightarrow \unM,\qquad i=0,1,p
\la{26}
\ee

and

\be
f_0{}^p:M_0\hookrightarrow M_p
\la{27}
\ee

with

\be
f_0=f_p\circ f_0{}^p\ .
\la{28}
\ee

For either of the fundamental strings the embedding condition

\be
E_{\aone}{}^{\una}=0
\la{29}
\ee

implies the equations of motion for the string. By using exactly the
same procedure as in the previous section we conclude that the embedding
condition will also hold for the D$p$-brane on which it ends, so that

\be
E_{\ap}{}^{\una}=0
\la{30}
\ee

In addition the standard embedding condition will also hold for the
worldvolume of the 0-brane boundary considered as a subsupermanifold of
$M_p$:

\be
E_{\a0}{}^{a_p}=0
\la{me2}
\ee

Each string has a Wess-Zumino 3-form $W_3=dZ_2$ which is simply the
pull-back of the NS 3-form 

\be
H_3=dB_2\ ,
\ee

on the target space, so we have

\be
W_3 = f_1^*H_3\ .
\la{31}
\ee

The Wess-Zumino form is exact on $M_1$ so

\be
W_3=dK_2
\la{32}
\ee

from which we conclude that

\be
K_2= dY_1 +f_1^*B_2 \qquad{\rm on} \ M_1\ .
\la{33}
\ee

On the boundary we identify $Y_1$ with the pull-back of a 1-form gauge
potential $A_1$ on $M_p$:

\be
Y_1=(f_1{}^p)^* A_1\ \qquad{\rm on} \ M_0\ .
\ee

We can thus identify the modified field strength 2-form ${\cF}_2$ on
$M_p$ as

\be
{\cF}_2=d A_1 +f_p^*B_2\ \qquad{\rm on} \ M_p\ ,
\la{34}
\ee

with the associated Bianchi identity

\be
d{\cF}_2= f_p^* H_3\ .
\ee

It follows from the constraints on the fundamental string that the only
non-vanishing component of the two-form $K_2$ is the one with purely
bosonic indices  and so it vanishes on the boundary. By a similar
argument to that given in the preceeding section, we therefore conclude
that

\be
{\cF}_{\ap B_p}=0\ .
\la{35}
\ee

The embedding condition \eq{30} and the ${\cF}$-constraint \eq{35} are
together sufficient to imply the equations of motion for the D$p$-brane
in all cases. They are not necessarily necessary, however. In type IIA
one can have $p=2,4,6,8$. For $p=2,4$, the embedding condition is
already enough to imply the equations of motion while for $p=6,8$, the
$\cF$-constraint is required as well. It is clear that an additional
constraint is required for $p=8$ since the brane has co-dimension one.
In this case the worldvolume multiplet determined by the embedding
condition is an entire scalar superfield. That an additional constraint
is required in the case of $p=6$ is less obvious since the worldvolume
multiplet is in this case a $d=7$ ``linear multiplet'', i.e. a
superfield whose leading component is three scalars and whose next
leading component is a spin-half field in seven dimensions. There is
also a 0-brane in type IIA for which the embedding condition alone is
sufficient to give the dynamics. In type IIB there are $D$-branes for
all odd $p$. For $p=1,3,5$ the embedding condition gives the equations
of motion while for $p=7,9$ the $\cF$-constraint is required as well.
For $p=7$ the worldvolume multiplet determined by the embedding
condition is a chiral scalar superfield, which is otherwise
unconstrained, and for $p=9$ there are no scalars so that it is clear
that an additional constraint is required. Further details of $D$-brane
embeddings will be found elsewhere \cite{hrss}. (The case of D$9$-brane
has been recently treated from the superembedding point of view in
\cite{abkz}).

\section{D-branes ending on D-branes}

In this section we shall consider D$p$-branes ending on D$(p+2)$-branes
of the Type IIA/B superstring theories in ten dimensions. The discussion
is very similar to the preceding two cases. It will enable us to recover
the results of \cite{chs} in a superspace approach.

The relevant supermanifolds are: the D$p$-brane manifold, $M_p$
(dimension $(p+1|16)$), its boundary $M_{p-1}=\del M_p$ (dimension
$(p|8))$, the worldvolume of the D$(p+2)$-brane $M_{p+2}$ (dimension
$(p+3|16)$) and the target space $\unM$ which is either type IIA or type
IIB superspace and which again has dimension $(10|32)$. The associated
embeddings are

\be
f_i:M_i\hookrightarrow \unM,\qquad i=p-1,p,p+2
\la{36}
\ee

and

\be
f_{p-1}{}^{p+2}:M_{p-1}\hookrightarrow M_{p+2}
\la{37}
\ee

with

\be
f_{p-1}=f_{p+1}\circ f_{p-1}{}^{p+2}\ .
\la{38}
\ee

For the D$p$-brane the embedding condition

\be
E_{\ap}{}^{\una}=0
\la{39}
\ee

is assumed to hold, as well as the $\cF$-constraint

\be
{\cF}_{\ap B_p}=0\ .
\la{40}
\ee

Using the same argument as before we deduce that the standard embedding
condition

\be
E_{\appt}{}^{\una}=0
\la{41}
\ee

will hold for the D$(p+2)$ brane as well. In addition the standard
embedding condition will also hold for the worldvolume of the $(p-1)$-brane
boundary considered as a subsupermanifold of $M_{p+2}$:

\be 
E_{\apmo}{}^{\appt} = 0 \la{me3} \ . 
\ee

The Wess-Zumino form for a D$p$-brane is $W_{p+2}=dZ_{p+1}$ where the
Wess-Zumino potential is given as the $p+1$-form component of an
inhomogeneous potential form 

\be
Z_{p+1}=(f_p^*C\,e^{\cF})_{p+1}\ ,
\la{42}
\ee

where $C$ is the sum of the RR potential forms on the target space. In
the case of massive type IIA D$p$-branes, the term $m f_p^* \o_{p+1}$,
where $m$ is the mass parameter and $\o_{p+1}(A,dA)$ is the Chern-Simons form,
has to be added to the right hand side \cite{bt}.

$W_{p+2}$ is a closed form on $M_p$ and by the same arguments that were
used before it must be exact, so that we can write

\be
W_{p+2}=d K_{p+1}
\la{43}
\ee

for some globally defined $(p+1)$-form $K_{p+1}$ on $M_p$. Clearly

\be
K_{p+1}=d Y_p + Z_{p+1}\ \qquad{\rm on} \ M_p\ .
\la{44}
\ee

On the boundary we identify $Y_p$ with the pull-back of a $p$-form
potential $A_p$ on $M_{p+2}$:

\be
Y_p=(f_{p-1}{}^{p+2})^* A_p\ \qquad{\rm on} \ M_{p-1}\ .
\ee

We also identify the 1-form potential of the D$p$-brane on the
boundary with the pull-back of the 1-form potential of the
D$(p+2)$-brane and use the same letter $A_1$. The field
strength $(p+1)$-form associated with $A_p$ is

\be
{\cF}_{p+1}=dA_p +  f^*_{p+2} (C\,e^{\cF})_{p+1}\ 
\qquad{\rm on} \ M_{p+2}\ ,
\la{45}
\ee

which obeys the Biachi identity  

\be
d {\cF}_{p+1}=(G e^{\cF})_{p+2}\ , \la{sbi}
\ee

with the definition

\be
G=dC-CH\ .
\ee

In the case of massive type IIA D$p$-branes, the Chern-Simons term $m
\o_{p+1}$ needs to be added to the right hand side of this definition
but the same Bianchi identity \eq{sbi} holds.

The equations for the $p$-brane imply that the only non-vanishing
component of $K_{p+1}$ is the one with purely bosonic indices and so we
deduce that the pull-back of ${\cF}_{p+1}$ to $M_{p-1}$ must vanish and
this implies the $\cF$-constraint for ${\cF}_{p+1}$, namely

\be
{\cF}_{\appt B_{p+2} C_{p+2}\ldots} =0\ .
\la{46}
\ee

The equations for the D$(p+2)$-brane derived from letting a D$p$-brane
end on it are therefore the standard embedding condition \eq{41}
together with ${\cF}$-constraints of the form of equation \eq{46} for
both a 2-form field strength ${\cF}_2$ and a $(p+1)$-form field strength
${\cF}_{p+1}$. These field strengths are essentially duals of one
another. To be more precise, at the linearised level ${\cF}_{ab}$ is the
dual of ${\cF}_{a_1\dots a_{p+1}}$, but in the full theory there are
non-linear corrections \cite{chs}.

\section{ Conclusions}

In this paper we have seen yet again the power of superembeddings in the
description of superbrane dynamics. Starting from simple geometrical
considerations having to do with the way an open M$2$-brane is embedded
in eleven dimensions or the way an open string or D$p$-brane is embedded
in ten dimensions, we were able to derive the superfield constraints
that govern the dynamics of the host branes on which these open branes
end. The constraints consist of the embedding condition of the host
brane and a constraint on a suitable field strength living on the host
brane, namely ${\cF}_3$ for the M$5$-brane, ${\cF}_2$ for the D$p$-brane
and ${\cF}_{p+1}$ for the D$(p+2)$ branes. The first two case are the
most familiar ones while the last case is somewhat novel in that it
contains a $p$-form potential $A_p$ as well as the usual Maxwell field
$A_1$, in accordance with the results of \cite{chs}. As mentioned
earlier, these are dual to each other (in a highly nonlinear fashion)
and consequently we expect to acquire new insights about duality
symmetries within this framework.

As a byproduct of our approach to open superbranes, we have obtained the
boundary embedding conditions \eq{18}, \eq{me2} and \eq{me3} that
characterize the embedding of the boundary manifolds within the host
branes, e.g. the closed string boundary of the M$2$-brane within the host
M$5$-brane, as given in \eq{18}. These constraints are not sufficient,
however, to put the boundary brane on-shell. 

Consider, for example, the case of D$p$-branes ending on
D$(p+2)$-branes, where there are boundary $(p-1)$-branes embedded in
$(p+2)$-branes. Note that $1\le p\le 7$. In the special case of the
D$7$-brane ending on the D$9$-brane, we expect the boundary brane to be
the familiar D$6$-brane moving in D$9$-brane, alias the ten dimensional
spacetime. Therefore let us concentrate on the remaining cases of
D$p$-branes ending on D$(p+2)$-branes with $1\le p\le 6$. All of these
are codimension $3$ embeddings in which the boundary D$(p-1)$-brane is
propagating in $(p+3)$-dimensions. The $3,4,5$ branes propagating in
$7,8,9$ dimensional target spacetimes, resepectively, have already been
encountered in the context of superembeddings in \cite{hs1}, where they
were called the L-branes. It was argued in \cite{hs1} that the
associated supermebedding constraints imply the equations appropriate to
linear multiplets which happen to be off-shell supermultiplets. 

Consider the L$5$-brane in $9$ dimensions (the description of the other
cases can be obtained by dimensional reduction). Its worldvolume
multiplet consists of three scalars, a $4$-form potential and a spinor
with $8$ real components. So the off-shell degrees of freedom count is
$3+5$ bosons and $8$ fermions. Interestingly, there are no auxiliary
fields in this multiplet. The existence of $4$-form potentials on the
worldvolume leads, by arguments similar to those of the previous
section, to the boundary ${\cF}$-constraint \cite{hrrs}

\be
d{\cF}_5= (f_5{}^8)^* G_6\ ,
\ee

where $G_6$ is a closed super-form in $9$ dimensions. However, the
system remains off-shell even in presence of this constraint. In order to
put the system on-shell, one has to construct an action that yields the
equations of motions. We will show elsewhere that this is indeed
possible \cite{hrrs}.

In passing we note that putting the linear multiplet on-shell means that
the $4$-form potential obeys a Maxwell type equation and therefore
on-shell it is dual to a scalar field. Since the fermions describe $4$
degrees of freedom, one then obtains a $4+4$ on-shell multiplet which is
essentially a hypermultiplet with one of the scalars dualized to a
$4$-form potential. This system is therefore intimately related to a
vertical reduction of a $5$-brane in $10$ dimensions, followed by the
dualization of the $10$th coordinate scalar to a $4$-form potential on
the $5$-brane worldvolume \cite{hs1}.

In this paper we focused on M$2$-brane ending on M$5$-brane, D$p$-branes
ending on D$(p+2)$-branes and fundamental string ending on D$p$-branes.
In the firt stwo cases we assumed that the open branes have single
component boundaries, while in the last case we let the two ends of the
open string lie on a single D$p$-brane, for simplicity. Not all of these
configurations are necessarily BPS saturated or anomaly free. While
anomaly freedom is essential, the BPS saturation is less crucial
property since the BPS states presumably constitute only a tiny fraction
of all possible states. 

The universal nature of the superembedding formalism suggests that it
can successfully be applied to many other generalizations of the systems
studied in this paper. For example, it can be applied straightforwardly
to D$p$-branes ending on NS $5$-branes. One can also treat
configurations in which the open M$2$-brane ends on an M$5$-brane at one
end and an M$9$-brane on the other, or both ends ending on M$9$-branes
(when we discuss M$9$-branes, we have in mind the Horawa-Witten picture
of such objects as boundaries of the eleven dimensional spacetime with
suitable topology \cite{hw1}).

One may also consider a system in which the open M$2$-brane has
multi-component boundaries which may end on any M$5$-branes or
M$9$-branes in all possible ways. It is clear that there is a rich
spectrum of possibilities due to the fact that the basic building blocks
can have a large class of nontrivial topologies. Further novel
possibilities can also arise because brane theories are intrinsically
nonlinear and consequently the topology of branes can change through
self-interactions. 

\bigskip

\noindent{\large \bf Acknowledments} 

It is a pleasure to thank M. Duff and P. Sundell for stimulating
discussions.

\ed